\documentclass[%
 reprint,
 amsmath,amssymb,
 aps,
prl,
]{revtex4-2}
\usepackage{amssymb}
\usepackage{amsthm}
\usepackage{graphicx}
\usepackage{dcolumn}
\usepackage{bm}

\allowdisplaybreaks

\begin{document}
\title{Information Thermodynamics of  Non-Hermitian Quantum Systems}

\author{Kui Cao${}^{1}$}
\author{Qian Du${}^{2}$}
\author{Su-Peng Kou${}^{1,}$}
\thanks{Corresponding author: spkou@bnu.edu.cn}

\affiliation{
    \textsuperscript{1}Center for Advanced Quantum Studies, Department of Physics, Beijing Normal University, Beijing 100875, China\\
    \textsuperscript{2}School of Physics and Electronic Engineering, Linyi University, Linyi 276000, China
}

\begin{abstract}
In this study, we uncover the intrinsic information processes in non-Hermitian quantum systems and their thermodynamic effects. We demonstrate that these systems can exhibit negative entropy production, making them potential candidates for information engines. We also identify a key informational quantity that can characterize phase transitions beyond the reach of traditional partition functions. This work enhances our understanding of the interplay between information and thermodynamics, providing a new perspective on non-Hermitian quantum systems.
\end{abstract}

\maketitle

The second law of thermodynamics, a fundamental principle in physics, states that the entropy production of a system is always non-negative. However, apparent exceptions to this law have been discussed, particularly in the context of Maxwell's demon. Investigations into these apparent anomalies have demonstrated that the perceived violations are resolved when accounting for the thermodynamic effects of information processing inherent in these systems\;\cite{Szilard1929,Szilard1964,Maruyama2009,1,2,3,4,5,6,7,9,JWang2021}. Studies of the thermodynamic effects of information processes are known as information thermodynamics. Excluding systems that directly implement feedback control, which evidently involve information processes, quantum systems involving information processes typically exhibit complexity characterized by intricate interactions\;\cite{8,fz1}.

Non-Hermitian quantum systems have garnered extensive research attention due to their diverse physical properties, such as $\mathcal{PT}$ symmetry breaking\;\cite{Bender98,Bender02,Bender05,Bender07}, the non-Hermitian skin effect\;\cite{Yao2018,Yao20182,Lee2019,Kunst2018,Yin2018, Yokomizo2019, KZhang2020,Slager2020, YYi, SMu2020,Okuma2021,Roccati,E.Lee, Shen
Mu, Tliu, bose mb1, bose mb2,CF2022}, and quantum phase transitions without gap closing\;\cite{Ueda2020,QWG2,QWG3}. Nonetheless, because non-Hermitian quantum systems are simply quantum systems with non-Hermitian Hamiltonians, they have often been considered too straightforward to exhibit non-trivial thermodynamic behavior. This presumed simplicity led earlier studies to assume that non-Hermitian systems follow the Boltzmann distribution, by directly substituting the Hermitian Hamiltonian with a non-Hermitian one in the density matrix formulation. As a result, these systems were believed to lack non-trivial information thermodynamics properties\;\cite{Quthe}. Contrary to this viewpoint, an alternative perspective suggests otherwise. Non-Hermitian systems can be conceptualized as open systems without quantum jump processes\;\cite{Yuto2018}. Choosing a null-jump trajectory from among various possible trajectories inherently involves information processes. Therefore, it is unreasonable to assume that these systems adhere to the Boltzmann distribution, which is related to the principle of maximum entropy. Recent research supports this assertion, demonstrating that non-Hermitian systems do not conform to the Boltzmann distribution\;\cite{note,DQ2022,Kcaoprr,cknhse,ckt}. This challenges earlier conclusions and suggests that non-Hermitian systems may exhibit unique information thermodynamic behaviors.

In this Letter, we demonstrate that non-Hermitian systems can exhibit behavior that appears to violate the second law of thermodynamics, analogous to Maxwell's demon effect. This observation suggests the potential for non-Hermitian systems to function as information engines that convert information into work. Specifically, we construct a two-level information engine to illustrate this concept. By extracting insights from the information thermodynamics framework, we identify a physical quantity that represents the requisite information for establishing a steady state in a non-Hermitian system. Using the Bosonic Hatano-Nelson model, we demonstrate that this quantity can characterize phase transitions driven by non-Hermiticity, which traditional partition functions fail to describe. Our work reveals a simple system with non-trivial information thermodynamics, enhancing our understanding of the fundamental role of information in thermodynamics. Additionally, it offers a novel perspective on non-Hermitian quantum physics, providing a valuable tool for exploring the intricate structure of non-Hermitian systems.

\textit{ Setup.}---In this Letter, we consider a non-Hermitian system coupled to a single Markovian thermal bath.  The time evolution equation of the  whole system's density matrix  $\rho_{t}$ is \cite{Yuto2018,BrodyGraefe2012,Kawabata2017,Fsong1}
\begin{equation}
\label{1}
i\frac{d}{dt}\rho_{t}= \hat{H}_{t}\rho_{t}- \rho_{t} \hat{H}_{t}^{\dag} + [ \mathrm{tr} (\hat{H}_{t}^\dag-\hat{H}_{t}) \rho_{t} ] \rho_{t}.
\end{equation}  
The initial density matrix $\rho_{t}$ is chosen to be a normalized Hermitian operator. In this equation, the first two terms on the right-hand side represent the standard Liouville evolution, while the remaining term ensures the normalization of the density matrix during evolution.    The Hamiltonian  
$\hat{H}_{t}$ comprises three components:
\begin{equation}
\hat{H}_{t}=\hat{H}_{\mathrm{NH}}\otimes \hat{I}_B+\hat{I}_S\otimes \hat{H}_{B}+\hat{H}_{BS}.
\end{equation}
$\hat{I}_S$ and $\hat{I}_B$ are the identity operators defined in the Hilbert spaces of the system and the heat bath, respectively. $\hat{H}_{B}$ is the Hamiltonian of the heat bath, $\hat
{H}_{BS}$ represents the coupling between system and the heat bath. We assume that $\hat{H}_{B}$ and $\hat{H}_{BS}$ are Hermitian.  $\hat{H}_{\mathrm{NH}}$ is the Hamiltonian of the system which is non-Hermitian.   We represent the Hamiltonian of the system as
\begin{equation}
\hat{H}_{\mathrm{NH}}=\hat{H}+i\hat{\Gamma}.
\end{equation}
Here, both $\hat{H}$ and $\hat{\Gamma}$ are Hermitian operators. $\hat{H}$ denotes the Hermitian component which plays a quintessential role in delineating the energy attributes of the system. $\hat{\Gamma}$ embodies the non-Hermitian part, serving to articulate the dissipation traits inherent to the system.

The above setting defines a time evolution equation for the reduced density matrix of the system  $\rho = \mathrm{tr}_B \rho_t$, which we denote as  
\begin{equation}
 \frac{d}{dt}\rho=\mathcal{D}\rho.
\end{equation}
The time evolution of $\rho$ conserves its trace while displaying non-linear characteristics.

We note that the setup of a non-Hermitian system coupled to a heat bath naturally emerges in scenarios where non-Hermitian systems are implemented. Previous studies have indicated that non-Hermitian quantum systems can be realized  through post-selection measurements on open systems \cite{Yuto2018,Ueda2020}. However, this represents an ideal scenario. In more realistic situations, the open system is also inevitably coupled to a thermal environment. Given this setup, one can derive that the effective time evolution equation for the entire system consists of a non-Hermitian system coupled to a Hermitian thermal bath. Details are provided in the supplementary material \cite{sm}.

The post-selection measurement inherent in non-Hermitian systems is a distinct information process.  In the next part, we will explore the thermodynamic effect resulting from this information process.

\textit{  Entropy production of non-Hermitian quantum systems.}---We derive the entropy production for the non-Hermitian system.
In thermodynamics, the entropy production $\dot{\Sigma}$ of a system is defined as
\begin{equation}\dot{\Sigma}\equiv \dot{S}-\beta\dot{Q},\end{equation}   where $\dot{S}$ is the rate of change of the entropy of the
system, $\dot{Q}$ is the heat current from the reservoir with inverse temperature  $\beta$. In the following part, we denote the rate of change of a physical quantity with respect to time by adding a dot over the quantity.



To derive the entropy production of the system, we need to clarify the definitions of entropy $S$ and heat $Q$. Given that a non-Hermitian system is an open system, its entropy can be defined as the von-Neumann entropy of the system's reduced density matrix:
\begin{equation}
S=-\mathrm{tr} \rho \ln \rho.  
\end{equation}

  To define $Q$, we first define the internal energy of a non-Hermitian system. 
We define the energy of the system as \begin{equation}U \equiv  Re(\mathrm{tr} \hat{H}_{\mathrm{NH}} \rho  ) = \mathrm{tr} \hat{H} \rho,\end{equation} where $Re(z)$ is the real part of complex number $z$. Under this definition,  the non-Hermitian term $\hat{\Gamma}$ only effects the ensemble of the system, not the energy of a given ensemble. Its expectation value corresponds to the average dissipation strength of the ensemble.

In analogy to the definitions of $Q$ and $W$ in Hermitian cases, the change in energy resulting from the Hamiltonian variation is defined as $W$, while the change in energy brought about by the state changes is defined as $Q$. As such, $Q$ and $W$ are respectively defined as 
\begin{align}
dQ\equiv  Re(\mathrm{tr}  \hat{H}_{\mathrm{NH}} d\rho  )=\mathrm{tr} \hat{H} d\rho, \nonumber \\   
dW\equiv  Re(\mathrm{tr} d\hat{H}_{\mathrm{NH}} \rho  )=\mathrm{tr} d\hat{H}  \rho.  
\end{align}
 These automatically satisfy the first law of thermodynamics and are consistent with  the definitions in Hermitian scenarios.

 We construct the inequality satisfied by the entropy production. We demonstrated that there exists an inequality\;\cite{sm}:
\begin{equation}
\mathrm{tr} (\ln \sigma-\ln \rho)\mathcal{D}\rho \geq Re \langle 2(\ln \sigma-\ln \rho)(\hat{\Gamma}- \langle \hat{\Gamma}\rangle) +  \Delta \rangle . \label{neq3}
\end{equation} 
Here, $ \langle \hat{A}\rangle \equiv \mathrm{tr} \hat{A} \rho $ represents the expected value of the operator $\hat{A}$ under the system's density matrix $\rho$.   $\Delta= \frac{d}{d\tau}  \ln (e^{i\hat{H} \tau} \sigma e^{-i\hat{H} \tau})|_{\tau=0}$.  $\sigma$ is the steady state which satisfy $\mathcal{D}\sigma=0$. The equality is taken at the system's steady state. Under the Hermitian limit, this inequality reduces to Spohn's inequality $\mathrm{tr} (\ln \sigma-\ln \rho)\mathcal{D}\rho \geq  0 $\;\cite{Spohn}.  

The steady state $\sigma$ of a typical non-Hermitian system can be obtained by solving the master equation or by utilizing the theoretical framework discussed in Ref.\;\cite{Kcaoprr,ckt}. Generally, $\sigma$ is not given by $\frac{e^{-\beta \hat{H}}}{\mathrm{tr}\,e^{-\beta \hat{H}}}$.  To better characterize this distinction, we define several physical quantities. We introduce a (Hermitian) thermal Hamiltonian $\hat{H}_{\mathrm{T}} \equiv -\frac{1}{\beta} (\ln \sigma + \ln \mathrm{tr} \, e^{-\beta \hat{H}})$, therefore the density matrix $\sigma$ can be regarded as the equilibrium density matrix of a system with Hamiltonian $\hat{H}_{\mathrm{T}}$ at an inverse temperature $\beta$. The term $\hat{V}_{\mathrm{NH}} \equiv \beta (\hat{H}_{\mathrm{T}} - \hat{H})$ characterizes the impact of non-Hermitian effects on the steady state. We refer to $\hat{V}_{\mathrm{NH}}$ as the non-Hermitian potential.

Employing the definitions of entropy and heat, together with the inequality in Eq.\;(\ref{neq3}), we obtain 
\begin{equation}
\label{123}
\dot{\Sigma}  \geq J_S,
\end{equation} 
here $J_S=\mathrm{tr} \hat{V}_{\mathrm{NH}} \dot{ \rho}+Re \langle 2(\ln \sigma-\ln \rho)(\hat{\Gamma}- \langle \hat{\Gamma}\rangle) +  \Delta \rangle$.  Inspired by past information flow theory, we
refer to $J_S$ as the information flow caused by
non-Hermiticity\;\cite{4,9}, which  characterizes the thermodynamic effects brought about by the information process. The first term of the information flow is reversible, while the second term is irreversible, originating from the decoherence and dissipation effects in non-Hermitian quantum systems.  For Hermitian system, there is obviously $J_S=0$, which conforms to the second law of thermodynamics. For non-Hermitian system, $J_S$ is sign-indefinite. Therefore the system can have negative entropy production.

%
%
%
%
%
%
%
%
%
%
%
%

During the system's transition between states 1 and 2, the total information flow  $I_S \equiv -\int_{1}^{2} dt J_S$ into the system can be expressed as
\begin{equation}
\label{124}
I_S=   {\mathcal{I}_{\mathrm{NH}}}|_1^2 -I_{diss}, 
\end{equation} 
where $I_{diss}=Re \int_{1}^{2} dt    \langle 2(\ln \sigma-\ln \rho)(\hat{\Gamma}- \langle \hat{\Gamma}\rangle) +  \Delta -\dot{\hat{V}}_{\mathrm{NH}} \rangle    $ which depending
on the specific process. ${\mathcal{I}_{\mathrm{NH}}}|_1^2\equiv({\mathcal{I}_{\mathrm{NH}}})_2-({\mathcal{I}_{\mathrm{NH}}})_1$. We refer to 
\begin{equation}
\mathcal{I}_{\mathrm{NH}}\equiv-\langle \hat{V}_{\mathrm{NH}}\rangle_{\rho}
\end{equation}
 as the non-Hermitian  information content depends only on the state of the
system.  We can prove that for steady state, $\mathcal{I}_{\mathrm{NH}}\geq 0$,
where the equal sign is obtained only when $\hat{V}_{\mathrm{NH}}=0$ \cite{sm}. Mathematically, $\mathcal{I}_{\mathrm{NH}}$
gives the part of information flow which does not depend on the process when
establishing the state $\rho$. It represents the intrinsic information cost necessary to establish the state $\rho$ and quantifies the extent to which information processes influence the formation of the steady state.   In the classical limit, $\mathcal{I}_{\mathrm{NH}}$ reduces to the minimum amount of classical information required to establish the steady state under suitable conditions \cite{sm}.   This physical quantity not only plays a role in studying thermodynamic processes but can also characterize phase and phase transitions of non-Hermitian systems, as shown below.

  \begin{figure}[ptb]
\includegraphics[clip,width=0.6\textwidth]{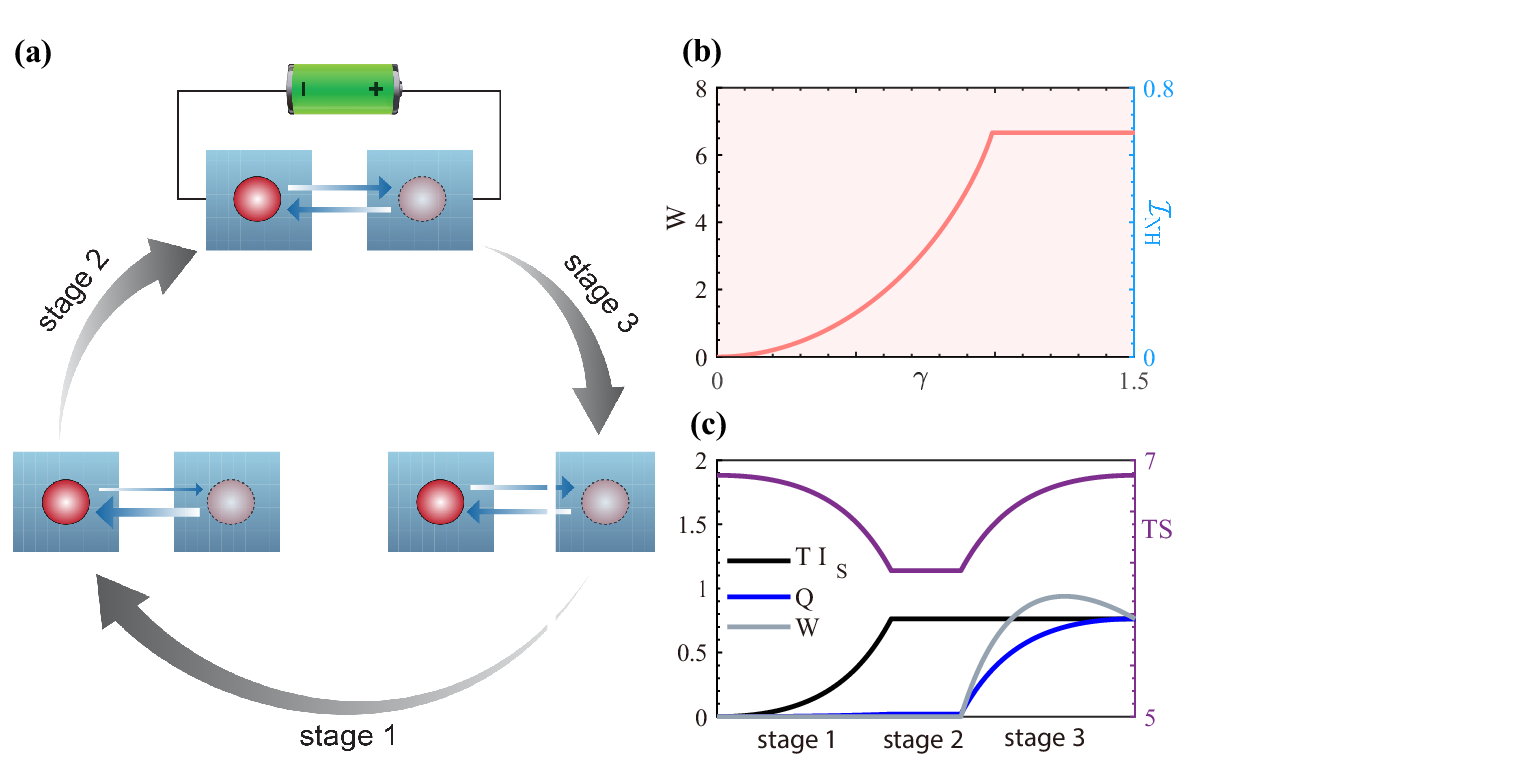}\caption{ (a) Illustration of the cycle for a two-level quantum heat engine. (b) Work done in the cycle with $T=10$. (c) Changes in various physical quantities during the cycle.  }
\label{fig1}
\end{figure}

\textit{Non-Hermitian information engine.}---The information thermodynamics of non-Hermitian systems can help us analyze the thermodynamic processes in non-Hermitian systems. We construct a intriguing heat engine that beyond the limits of the Kelvin-Planck statement of the second law---it can absorb heat and perform work  from  a single heat reservoir in a cycle.   Our heat engine is a non-Hermitian system coupled to a single heat bath with temperature $T$. Its cycle is divided into three stages:

\textit{Stage 1}: Starting with a Hermitian system with Hamiltonian $\hat
{H}$.
Adjust a real number $\gamma_s \in [0,1]$ slowly to make the Hamiltonian of the system  from $\hat{H}$ to
$\hat{H}_{\mathrm{NH}}\equiv \hat{H}+i\hat{\Gamma}$ along $\hat{H}_{\mathrm{NH}}(\gamma_s)\equiv \hat{H}+i\gamma_s \hat{\Gamma}$.

\textit{Stage 2}: Quickly changed the system's Hamiltonian to $\hat
{H}_{\mathrm{1}}\equiv  \hat{H}_{\mathrm{T}}+ E_1$. Here $\hat{H}_{\mathrm{T}}$ is the thermal Hamiltonian of final state of Stage 1, $E_1$ is chosen as a real number to keep the energy unchanged at this stage.

\textit{Stage 3}: Slowly adjust the system to make the Hamiltonian from $\hat{H}_{\mathrm{1}}$ back to $\hat{H}$.

Since the system remains near steady state throughout the entire cycle, the  work done by the system in the cycle  can be easily obtained. We have $W = -\oint_{cycle} dt (\dot{U}- T\dot{S}+ TJ_{S})= - T\int_{stage \ 1} dt J_{S}$.   
After   straightforward calculation,  it is found that the total information flow in this step is just equal to the final state's non-Hermitian  information content. Therefore, we have
\begin{equation}
W= T\cdot \mathcal{I}_{\mathrm{NH}}.\label{e1}%
\end{equation}

For illustration, we consider a two-level quantum heat engine. Its Hamiltonian is
\begin{equation}
\hat{H}_{\mathrm{NH}}= \sigma_x + i\gamma {\sigma_y},\label{m1}
\end{equation}
 here $\sigma_{i} \ (i=x,y,z)$ is the Pauli matrix, and which base is  $\{ \left \vert 1\right \rangle ,\left \vert 2\right \rangle \}$.  The real number $\gamma$
controls the nonreciprocity between  states $\left \vert 1\right \rangle $ and $\left \vert 2\right \rangle $. The coupling between the system and the heat bath is set as $\hat{H}_{BS}=  \lambda_{1} \left \vert 1\right \rangle
 \left \langle 1\right \vert\otimes \hat{B}_{1}+\lambda_{2} \left \vert 2\right \rangle
 \left \langle 2\right \vert\otimes \hat{B}_{2}$, where $\hat{B}_{1,2}$ is an arbitrary Hermitian operator defined on the Hilbert space of the heat bath and $\lambda_{1,2}$ is a real number much less than 1.
 Through straightforward calculations, we achieve $\hat{V}_{\mathrm{NH}}\sim -\gamma  \sigma_z $, which indicates the non-Hermitian effects makes particles tend to stay at a certain point\;\cite{DQ2022,Kcaoprr}.   By calculating the expected value of $\hat{V}_{\mathrm{NH}} $ at the steady state, we get 
the $\mathcal{I}_{\mathrm{NH}}$, as shown in Fig.\;\ref{fig1}(b). In particular, at the high-temperature limit $T \gg 1$, $\mathcal{I}_{\mathrm{NH}}$ has a simple expression:  \begin{equation}
\mathcal{I}_{\mathrm{NH}}= 
\begin{cases} 
\frac{1}{2}[ (1+\gamma) \ln (1+\gamma)+(1-\gamma) \ln (1-\gamma)], & \text{if } |\gamma|\leq 1,\\
\ln2, & \text{if } |\gamma| \geq 1.
\end{cases}
\end{equation}
 Obviously, this result violates the Kelvin-Planck statement of the second law of thermodynamics.   We also calculated the changes of various physical quantities in each stage, as shown in Fig.\;\ref{fig1}(c). The non-Hermitian effect played an important  role in stage 1. It inputs information flow to the system, decreasing the entropy of the system without affecting the energy. The subsequent stage \textquotedblleft convert\textquotedblright \ this information into work.

The first derivative of  $\mathcal{I}_{\mathrm{NH}}$ has a mutation at $|\gamma|=1$. It corresponds to the exceptional point (EP) of the system, where the energy spectrum of the system is transformed from real to complex\;\cite{Bender98,Bender02,Bender05,Bender07}. The informational perspective gives another understanding of EP---it is the critical point from the state under partial control to the state under complete control by the Maxwell's demon generated by information  processes.
%
%

This result illustrates that the non-Hermitian  information content not only serve in characterizing the thermodynamic processes of the non-Hermitian system, but also play an instrumental role in understanding the phases and phase transitions in non-Hermitian systems. The phase transitions considered here involve changes in the energy spectrum, which can be captured by the partition function $Z=\mathrm{tr}\exp(-\beta \hat{H}_{\mathrm{NH}})$. In the following part, we will demonstrate that this information quantity can also reveal phase transitions that conventional partition functions fail to detect.

\textit{Informational description of non-Hermitian system.}---Informational quantities provide an alternative perspective for understanding the phases and phase transitions in non-Hermitian systems. We take the Bosonic Hatano-Nelson model with open boundary conditions at finite temperature as an example. This model was studied in Ref.\;\cite{cknhse}.
It is a representative model with non-Hermitian skin effect. The Hamiltonian of the model is\;\cite{Hatano Nelson}
\begin{equation}
\hat{H}_{\mathrm{HN}}=J\sum_{j}^{L-1}(e^{-\frac{g}{2L}%
}a_{j+1}^{\dagger}a_{j}+e^{\frac{g}{2L}}a_{j}^{\dagger
}a_{j+1}),\label{HN}%
\end{equation}
where $L$ is the number of lattice sites, $J$ is the hopping parameter and the
real number $g$  controls the overall skin effect strength of the system. $a_{j}^{\dag}/a_{j}$ is the
create/annihilate operator for particle at $j$-th lattice. We have
$[a_{i},a_{j}^{\dagger}]=\delta_{ij}$. The energy spectrum of the model is $E_k=2J\cos
\frac{2\pi k}{L}$, $k=1,2,3,...,L$, which does not related to $g$. For simplicity, we assume that the total particle number
$N_{B}$ is equal to the number of lattice sites, i.e., $N_{B}=L$, and consider the thermodynamic limit $L\rightarrow \infty$.

Up to a constant, the  
 thermal Hamiltonian $\hat{H}_{\mathrm{T}}$ is obtain as\;\cite{cknhse}

\begin{equation}
\hat{H}_{\mathrm{T}}=\sum_{j}J(a_{j+1}^{\dagger}a_{j}%
+a_{j}^{\dagger}a_{j+1})+\sum_{j} V%
(a_{j}^{\dagger}a_{j}\cdot j/L),
\end{equation}
where  $V=     g T $. $\hat{H}_{\mathrm{T}}$ can be expressed as the Hamiltonian of reciprocal model $\hat{H} $ plus the linear field with potential $V$.
In the Ref.\;\cite{cknhse}, the linear field was interpreted as an electric field effect. We now confirm that it corresponds very well to the Maxwell's pressure demon effect; see Supplemental Material. 

\begin{figure}[ptb]
\includegraphics[width=9cm ]{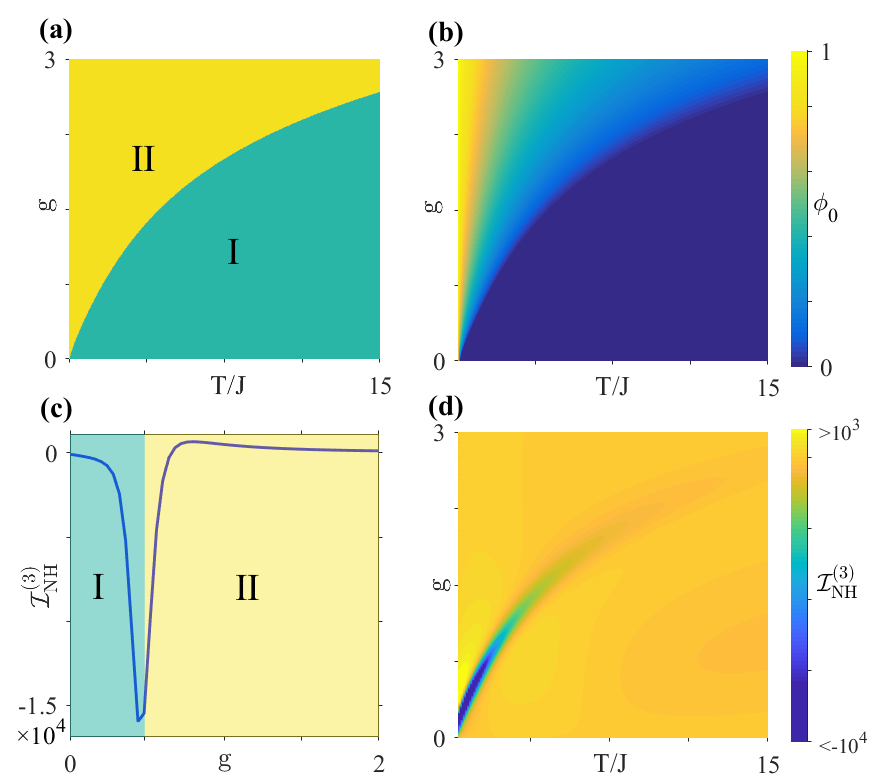}\caption{ (a) Phase diagram for the 1D many-body Bosonic Hatano-Nelson model: yellow represents the normal phase, and green indicates the phase with spontaneous $U(1)$ symmetry breaking without a Goldstone mode. (b) Order parameter $\phi_{0}$ for the  Hatano-Nelson model. (c) Third derivative of non-Hermitian information content for the  non-Hermitian model with $L=5000$ and $T=1$. (d) Third derivative of non-Hermitian information content for the   Hatano-Nelson model with $L=5000$. }%
\label{fig2}
\end{figure}

This model has a $U(1)$ symmetry, i.e.,  its  Hamiltonian invariant under the
global gauge transformation $a_{j}^{\dag}\rightarrow e^{-i\theta}a_{j}^{\dag
},a_{j}\rightarrow e^{i\theta}a_{j}$. To characterize the phase transition, we
define an order parameter $\phi_{0}=|\mathrm{tr}\ a_{0}\sigma%
|^{2}=N_{0}/N_{B}$, where $a_{0}$ is the
annihilation operator for the ground state of thermal Hamiltonian $\hat{H}_{\mathrm{T}}$, and $N_{0}$ is the number of particles on the state. When a macroscopic number of particles condense on
the state,  we have $\phi_{0}\neq0$, at which the phase transition
occurs. On the other hand, the order parameter is zero if the state has $U(1)$
symmetry.  Therefore, the
system exhibits spontaneous $U(1)$ symmetry breaking when this phase
transition occurs.
 We estimate the values of critical points for the phase transition. The
number of particles $N_{0}$ of the thermal \textquotedblleft ground
state\textquotedblright \  is $N_{0}=N_{B}-\int
_{e_{0}^{\mathrm{T}}}^{\infty}de^{\text{\textrm{T}}}\frac{1}{e^{\beta
 (e^{\text{\textrm{T}}}-e_{0}^{\mathrm{T}})}-1}\rho
(e^{\text{\textrm{T}}})$, where $e^{\mathrm{T}}$ is the thermal Hamiltonian's
energy level, $e_{0}^{\mathrm{T}}$ is the energy level with the lowest energy, and $\rho(e^{\text{\textrm{T}}})$ are the
density of states derived from $\hat{H}_{\mathrm{T}}$. Therefore, the phase
transition occurs at $N_{B}=\int_{e_{0}^{\mathrm{T}}}^{\infty}%
de^{\text{\textrm{T}}}\frac{1}{e^{\beta (e^{\text{\textrm{T}}%
}-e_{0}^{\mathrm{T}})}-1}\rho(e^{\text{\textrm{T}}})$. The values of critical
points can be obtained via straightforward calculations. At low temperature
$ T\ll J$, we have $g_{c}\sim \frac{ T}{J}$ or
$ T_{c}\sim Jg;$ and at high temperature
$ T\gg J$, we have $g_{c}\sim \ln \frac{ T}{J}$
or $ T_{c}\sim Je^g.$ The phase transition
occurs in the region of strong nonreciprocal strength  or low
temperature, as seen in Fig.\;\ref{fig2}(a). It can be regarded as Bose-Einstein condensate (BEC) with
particles condense on the ground state of thermal Hamiltonian. 

This phase transition is also unique in that it violates the Goldstone theorem—it results in symmetry breaking without the emergence of Goldstone modes. This anomaly occurs because the phase transition can be induced by adjusting the non-reciprocity of the hoppings without altering the energy spectrum.  This also making it undetectable by partition functions $Z=\mathrm{tr}\exp (-\beta \hat{H}_{\mathrm{NH}})$. However, it can be revealed by non-Hermitian information content $\mathcal{I}_{\mathrm{NH}}$. To show this, we calculate the
  $\mathcal{I}_{\mathrm{NH}}$:
\begin{equation}
\mathcal{I}_{\mathrm{NH}}=\sum_{j,n}\frac{-gj|\langle
j|e_{n}^{\mathrm{T}}\rangle|^{2}/L}{e^{{\beta}(e_{n}^{\mathrm{T}}%
-\mu)}-1}+\sum_{n}\ln \frac{1-e^{\beta_{ }(e_{n}-\mu_{0})}}%
{1-e^{{\beta}(e_{n}^{\mathrm{T}}-\mu)}}+L(\mu-\mu_{0}),
\end{equation}
where  $e_{n}$, $\mu_{0}$ is the single-body energy level and chemical
potential for $g=0$ case. There is singularity in the expression of $\mathcal{I}_{\mathrm{NH}}$,
in which the location is the same as the previous critical point, see Fig.\;\ref{fig2}. This fact indicates that phase transition can be regarded as a result of the competition between \textquotedblleft energy-dominated\textquotedblright and the \textquotedblleft  information-dominated\textquotedblright physics. Furthermore, when  $g$ is large, the non-Hermitian informational content of the system is significantly high at elevated temperatures. This explains why the system can maintain the BEC phase under such conditions: achieving this state requires injecting substantial information into the system during its realization, which in turn reduces the system's entropy.


\textit{Conclusions.}---In this work, we examined the information processes in non-Hermitian systems and their thermodynamic effects. Our findings show that these processes can  input information flow, allowing non-Hermitian systems to apparently bypass the second law of thermodynamics. This finding posits an intriguing parallel, where non-Hermitian effects can be perceived as a form of Maxwell's demon effect. For example, the non-Hermitian skin effect can be interpreted as Maxwell's pressure demon effect. We also found that the total information flow in a process is linked to a state variable, representing the essential information needed to achieve steady state in a non-Hermitian system. This state variable captures phase transitions induced by non-Hermitian characteristics, which cannot be described by traditional partition functions. Our research thus establishes a theoretical framework for analyzing thermal processes in non-Hermitian systems and provides a novel approach to understanding their phases and phase transitions. We believe our work will deepen the understanding of the interplay between information and thermodynamics and pave the way for exploring exotic quantum phases and transitions in non-Hermitian systems.

 \acknowledgments This work was supported by the Natural Science Foundation of China (Grant No. 12174030) and National Key R$\&$D Program of China (Grant No. 2023YFA1406704).



\begin{thebibliography}{99}                                                                                               %



\bibitem {Szilard1929} L. Szilard, Über die Entropieverminderung in einem thermodynamischen System bei Eingriffen intelligenter Wesen, Z. Phys. \textbf{53}, 840-856 (1929).


\bibitem {Szilard1964}L. Szilard, On the decrease of entropy in a thermodynamic system by the intervention of intelligent beings, Behav. Sci. \textbf{9}, 301-310 (1964).



\bibitem{1}D. Mandal, and C. Jarzynski, Work and information processing in a solvable model of Maxwell's demon, Proc. Natl. Acad. Sci. \textbf{109}, 11641-11645 (2012).

\bibitem{2}P. Strasberg, G. Schaller, T. Brandes and M. Esposito,  Thermodynamics of a physical model implementing a Maxwell demon, Phys. Rev. Lett. \textbf{110}, 040601 (2013).

\bibitem{3}T. Sagawa,  and M. Ueda,  Generalized Jarzynski equality under nonequilibrium feedback control, Phys. Rev. Lett. \textbf{104}, 090602 (2010).

\bibitem{4}J. M. Horowitz,  and M. Esposito,  Thermodynamics with continuous information flow,  Phys. Rev. X  \textbf{4}, 031015 (2014).

\bibitem {Maruyama2009}K. Maruyama , F. Nori,  and V. Vedral,  Colloquium: The physics of Maxwell's demon and information, \textit{Rev. Mod. Phys.} \textbf{81}, 1 (2009).

\bibitem{5}J. M. R. Parrondo, J. M. Horowitz,  and T. Sagawa,  Thermodynamics of information,  Nat. Phys.  \textbf{11}, 131-139 (2015).

\bibitem{6}J. V. Koski, A. Kutvonen, I. M. Khaymovich, T. Ala-Nissila  and J. P. Pekola,  On-chip Maxwell's demon as an information-powered refrigerator, Phys. Rev. Lett. \textbf{115}, 260602 (2015).

\bibitem{7}N. Cottet, S. Jezouin, L. Bretheau, P. Campagne-Ibarcq, Q. Ficheux,  J. Anders, A. Auffèves, R. Azouit , P. Rouchon  and   B.  Huard, Observing a quantum Maxwell demon at work,  Proc. Natl. Acad. Sci.  \textbf{114}, 7561-7564 (2017).





\bibitem{9} K. Ptaszyński and M. Esposito,  Thermodynamics of quantum information flows, Phys. Rev. Lett. \textbf{122}, 150603 (2019).

\bibitem {JWang2021} Q. Zeng  and J. Wang,  New fluctuation theorems on Maxwell's demon,  Sci. Adv.  \textbf{7}, eabf1807 (2021).




\bibitem{8} K. Ptaszyński, Autonomous quantum Maxwell's demon based on two exchange-coupled quantum dots,  Phys. Rev. E  \textbf{97}, 012116 (2018).

\bibitem {fz1}P. Strasberg, G. Schaller, T. L. Schmidt, and M. Esposito, Fermionic reaction coordinates and their application to an autonomous Maxwell demon in the strong coupling regime, Phys. Rev. B \textbf{97}, 205405 (2018).





\bibitem {Bender98} C. M. Bender  and  S. Boettcher, Real Spectra in Non-Hermitian Hamiltonians Having PT Symmetry, Phys. Rev. Lett. \textbf{80}, 5243-5246 (1998).

\bibitem {Bender02} C. M. Bender,  D. C. Brody  and H. F. Jones,  Complex extension of quantum mechanics, Phys. Rev. Lett. \textbf{89}, 270401 (2002).

\bibitem {Bender05}C. M. Bender,  Introduction to PT-symmetric quantum theory,  Contemp. Phys.  \textbf{46}, 277-292 (2005).

\bibitem {Bender07}C. M. Bender, Making sense of non-Hermitian Hamiltonians, Rep. Prog. Phys. \textbf{70}, 947 (2007).

%
%
%
%
%
%


\bibitem {Yao2018} S. Yao  and Z.  Wang, Edge states and topological invariants of non-Hermitian systems, Phys. Rev. Lett. \textbf{121}, 086803 (2018).


\bibitem {Yao20182}S. Yao,   F. Song  and  Z. Wang, Non-Hermitian Chern bands, Phys. Rev. Lett. \textbf{121}, 136802 (2018).

\bibitem {Kunst2018}  F. K. Kunst,  E. Edvardsson,  J. C. Budich  and  E. J. Bergholtz, Biorthogonal bulk-boundary correspondence in non-Hermitian systems, Phys. Rev. Lett. \textbf{121}, 026808 (2018).


\bibitem {Yin2018}C. Yin, H. Jiang,  L.  Li,  R. Lü  and  S. Chen, Geometrical meaning of winding number and its characterization of topological phases in one-dimensional chiral non-Hermitian systems,  Phys. Rev. A  \textbf{97}, 052115 (2018).







\bibitem {Lee2019}C. H. Lee  and R. Thomale,  Anatomy of skin modes and topology in non-Hermitian systems,  Phys. Rev. B \textbf{99}, 201103(R) (2019).


\bibitem {Yokomizo2019}K. Yokomizo   and S. Murakami,  Non-bloch band theory of non-hermitian systems, Phys. Rev. Lett. \textbf{123}, 066404 (2019).



\bibitem {KZhang2020} K. Zhang,  Z. Yang  and C. Fang,  Correspondence between winding numbers and skin modes in non-Hermitian systems, Phys. Rev. Lett. \textbf{125}, 126402 (2020).

\bibitem {Slager2020} D. S. Borgnia,  A. J. Kruchkov  and R.-J. Slager,  Non-Hermitian boundary modes and topology, Phys. Rev. Lett. \textbf{124}, 056802 (2020).

\bibitem {YYi}Y. Yi   and Z. Yang,  Non-Hermitian skin modes induced by on-site dissipations and chiral tunneling effect, Phys. Rev. Lett. \textbf{125}, 186802 (2020).

\bibitem {SMu2020} L. Li,  C. H.  Lee, S. Mu  and J.  Gong,  Critical non-Hermitian skin effect, Nat. Commun.  \textbf{11}, 5491 (2020).




\bibitem {Okuma2021} N. Okuma and  M. Sato, Non-hermitian skin effects in hermitian correlated or disordered systems: Quantities sensitive or insensitive to boundary effects and pseudo-quantum-number, Phys. Rev. Lett. \textbf{126}, 176601 (2021).


\bibitem {Roccati}F. Roccati,  Non-Hermitian skin effect as an impurity problem,  Phys. Rev. A  \textbf{104}, 022215 (2021).

\bibitem{CF2022} K. Zhang, Z. Yang   and C. Fang,  Universal non-Hermitian skin effect in two and higher dimensions, Nat. Commun.  \textbf{13}, 2496 (2022).

\bibitem {E.Lee}E. Lee, H. Lee  and B.-J. Yang,  Many-body approach to non-Hermitian physics in fermionic systems,  Phys. Rev. B \textbf{101}, 121109(R) (2020).

\bibitem {Shen Mu}S. Mu,  C. H.  Lee, L. Li   and  J. Gong, Emergent Fermi surface in a many-body non-Hermitian fermionic chain,  Phys. Rev. B \textbf{102}, 081115(R) (2020).



\bibitem {Tliu}T. Liu,  J. J. He, T. Yoshida, Z.-L. Xiang  and  F. Nori, Non-Hermitian topological Mott insulators in one-dimensional fermionic superlattices,  Phys. Rev. B \textbf{102}, 235151 (2020).



\bibitem {bose mb1} D.-W. Zhang, Y.-L. Chen,  G.-Q. Zhang,  L.-J. Lang, Z. Li  and  S.-L. Zhu, Skin superfluid, topological Mott insulators, and asymmetric dynamics in an interacting non-Hermitian Aubry-André-Harper model,  Phys. Rev. B \textbf{101}, 235150 (2020).

\bibitem {bose mb2}Z. Xu  and S. Chen,  Topological Bose-Mott insulators in one-dimensional non-Hermitian superlattices,  Phys. Rev. B \textbf{102}, 035153 (2020).





\bibitem {Ueda2020}N. Matsumoto, K. Kawabata, Y. Ashida, S. Furukawa and M. Ueda, Continuous Phase Transition without Gap Closing in Non-Hermitian Quantum Many-Body Systems,  Phys. Rev. Lett. \textbf{125}, 260601 (2020).

\bibitem {QWG2} F. Yang, H. Wang, M.-L. Yang, C.-X. Guo, X.-R. Wang, G.-Y. Sun and S.-P. Kou, Hidden continuous quantum phase transition without gap closing in non-Hermitian transverse Ising model, New J. Phys. \textbf{24}, 043046 (2022).

\bibitem {QWG3} S. Sayyad and J. L. Lado, Topological phase diagrams of exactly solvable non-Hermitian interacting Kitaev chains, Phys. Rev. Res. \textbf{5}, L022046 (2023).

 \bibitem {Quthe} B. Gardas, S. Deffner  and  A. Saxena, Non-hermitian quantum thermodynamics,  Sci. Rep. \textbf{6}, 23408 (2016).

\bibitem {Yuto2018} Y. Ashida and  M. Ueda, Full-counting many-particle dynamics: Nonlocal and chiral propagation of correlations, Phys. Rev. Lett. \textbf{120}, 185301 (2018).
 
\bibitem {note}A significant issue remains in the assumption of early studies. In the context of standard projective measurements, the density matrix should be Hermitian. However, the Boltzmann distribution described here yields a non-Hermitian density matrix.

\bibitem {DQ2022} Q. Du, K.  Cao  and  S.-P. Kou, Physics of PT-symmetric quantum systems at finite temperatures,  Phys. Rev. A  \textbf{106}, 032206 (2022).

\bibitem{Kcaoprr} K. Cao and S.-P. Kou, Statistical mechanics for non-Hermitian quantum systems, 
Phys. Rev. Res. \textbf{5}, 033196 (2023).

\bibitem {cknhse} K. Cao, Q. Du, and S.-P Kou, Many-body non-Hermitian skin effect at finite temperatures, Phys. Rev. B \textbf{108}, 165420 (2023).
\bibitem {ckt} K. Cao and S.-P. Kou, Topological phases of many-body non-Hermitian systems, Phys. Rev. B \textbf{109}, 155434 (2024).









\bibitem{BrodyGraefe2012}D. C. Brody and E.-M. Graefe,  {Mixed-State Evolution in the Presence of Gain and Loss},  Phys. Rev. Lett. \textbf{109}, 230405 (2012).

\bibitem{Kawabata2017}K. Kawabata, Y. Ashida, and M. Ueda,  {Information Retrieval and Criticality in Parity-Time-Symmetric Systems}, Phys. Rev. Lett. \textbf{119}, 190401 (2017).



\bibitem {Fsong1} K. L. Zhang  and  Z. Song,  Quantum phase transition in a quantum Ising chain at nonzero temperatures, Phys. Rev. Lett. \textbf{126}, 116401 (2021).

\bibitem{sm} See Supplemental Material.

\bibitem {Spohn}H. Spohn,  Entropy Production for Quantum Dynamical Semigroups,  J. Math. Phys. (N.Y.)  \textbf{19}, 1227 (1978).

 

\bibitem {Hatano Nelson}N. Hatano and D. R. Nelson,  Localization Transitions in Non-Hermitian Quantum Mechanics, Phys. Rev. Lett. \textbf{77}, 570-573 (1996).









 










\end{thebibliography}
\end{document}